\documentclass[a4paper, 12pt, english]{article}

\usepackage[utf8]{inputenc}
\usepackage{amsmath,amssymb}
\usepackage{graphicx}
\usepackage{subfig}
\usepackage[colorinlistoftodos]{todonotes}

\usepackage{indentfirst}
\usepackage{verbatim}
\usepackage{textcomp}
\usepackage{gensymb}

\usepackage{relsize}

\usepackage{lipsum}% http://ctan.org/pkg/lipsum
\usepackage{xcolor}% http://ctan.org/pkg/xcolor
\usepackage{xparse}% http://ctan.org/pkg/xparse
\NewDocumentCommand{\myrule}{O{1pt} O{2pt} O{black}}{%
  \par\nobreak % don't break a page here
  \kern\the\prevdepth % don't take into account the depth of the preceding line
  \kern#2 % space before the rule
  {\color{#3}\hrule height #1 width\hsize} % the rule
  \kern#2 % space after the rule
  \nointerlineskip % no additional space after the rule
}
\usepackage[section]{placeins}

\usepackage{booktabs}
\usepackage{colortbl}%

\usepackage[obeyspaces]{url}
\usepackage{etoolbox}
\usepackage[colorlinks,citecolor=black,urlcolor=blue,bookmarks=false,hypertexnames=true]{hyperref} 

\usepackage{geometry}
\begin{document}

%************************************TITLE PAGE**************************************%
\begin{titlepage}
\begin{center}

\includegraphics[width=0.5\textwidth]{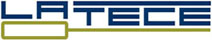}\\
\textbf{\large Laboratoire de Recherches sur les Technologies du Commerce Électronique}\\[0.2cm]

\vspace{10pt}

\includegraphics[width=0.5\textwidth]{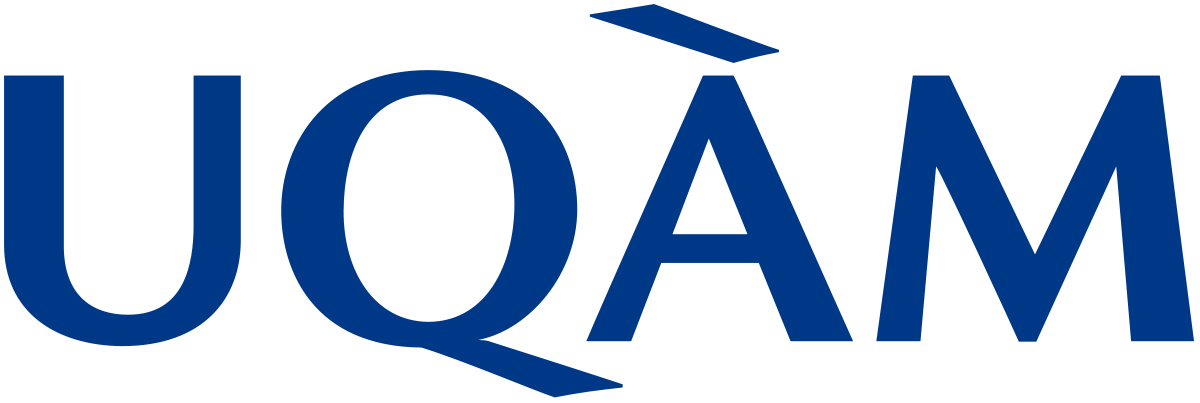}\\[1cm]

\textbf{\LARGE Universit\'e du Qu\'ebec \`a Montr\'eal}\\[0.5cm] 
\vspace{20pt}

\par
\vspace{20pt}
\myrule[1pt][7pt]
\textbf{\Large  What Should You Know Before Developing a Service Identification Approach}\\
\myrule[1pt][7pt]

\vspace{35pt}
\large{\textbf{Anas Shatnawi, Hafedh Mili, Manel Abdellatif, \\Ghizlane El Boussaidi, Yann-Ga\"el Gu\'eh\'eneuc, \\Naouel Moha, Jean Privat}}\\[0.3cm]

\textit{LATECE Technical Report 2017-2, LATECE Laboratoire, Universit\'e du Qu\'ebec \`a Montr\'eal, Canada}

\vspace{200pt}
\small{July, 2017}
\end{center}

\par
\vfill
\begin{center}
\end{center}
\end{titlepage}

%************************************TABLE OF CONTENTS**************************************%

%  %Sumário
%  \newpage
%  \tableofcontents
%  \thispagestyle{empty}
%  %End Sumário

%********************************%
%***********SECTION 1************%
%********************************%
\newpage

\begin{center}
\textbf{\Large  What Should You Know Before Developing a Service Identification Approach} \\[0.9cm]

{Anas Shatnawi\footnote{anasshatnawi@gmail.com}, Hafedh Mili\footnote{mili.hafedh@uqam.ca}, Manel Abdellatif, Ghizlane El Boussaidi, \\Yann-Ga\"el Gu\'eh\'eneuc, Naouel Moha, Jean Privat}\\[0.5cm]

\textit{LATECE Technical Report 2017-2, LATECE Laboratoire, Universit\'e du Qu\'ebec \`a Montr\'eal, Canada}
\end{center}

\begin{abstract}
In this paper, we answer a set of research questions that are required to develop service identification approach based on the analysis of object-oriented software. Such research questions are: 
    (1) what is a service,
    (2) how are services different from software components, 
    (3) what are types of services,
    (4) what are existing service identification approaches that consider service types into account, and 
    (5) how to identify services based on the object-oriented source code with respect to their types.
    Our methodology is based on performing a literature review to identify the answers of these research questions. Also, we propose a taxonomy of service types.
\end{abstract}

\section{Introduction}
\label{Introduction}
Legacy software are software that have been developed based on \textit{outdated} technologies, however they still give \textit{significant values} to the enterprises \cite{sneed2006integrating}. 
Besides their well-known advantages, legacy software still suffer from several drawbacks: maintenance cost, scalability, portability, etc. 
Due to the knowledge embedded in these legacy software, enterprises cannot easily \textit{replaced} such software. Therefore, enterprises need to migrate their legacy software to \textit{more loosely coupled architectures} such as \textit{Service-Oriented Architectures} (SOA). 

The migration of such software to SOA does not only allow enterprises to \textit{invest the values} of legacy software, but it also enables \textit{the integration with new advanced technologies}. i.e., services can be shared and (re)used as \textit{stand-alone functionalities} through
their provided and required interfaces, and can be successfully
\textit{deployed in a cloud environment}.

Legacy software migration to SOA involves two processes: (i) service identification and (ii) service packaging. Service identification aims to reverse engineer clusters of useful functionalities that can be good candidates for services. It was the goal of several approaches like \cite{nakamura2009extracting} \cite{alahmari2010service} \cite{fuhr2011using} \cite{grieger2014architectural} \cite{adjoyanservice} \cite{zhang2005service}. Service packaging aims to repackage these clusters in the "format" of a SOA model. This has been supported by many approaches such as \cite{alshara2016materializing}.

After analyzing the literature of service identification process, we identify that existing approaches do not have a clear idea about several important issues. Such issues are, but not limited to what is a service and how does it differ from a software component, taxonomies of service types. 
Therefore, in this paper, we identify-and solve- a set of five research questions that are related to these issues. These research questions are:

\begin{enumerate}
\item What is a service? 

\item How are services different from components?

\item {What are the different service types?}

\item {What are the existing service identification approaches that consider service types into account?}

\item {How to identify services based on the analysis of legacy source code with respect to service types?}
\end{enumerate}

Our methodology to answer these research question is based on analyzing existing definitions of services, analyzing existing service taxonomies that define the different types of services, building a service taxonomy of service types as a result of the analysis of the existing ones, analyzing existing service identification approaches that considered different service types into account during the identification process, and
providing criteria that can be used to identify services within legacy applications based on service types.

The rest of this report is organized as follows. We discuss service definitions in Section \ref{sec:what-services}. Then, Section \ref{sec:service-vs-component} differentiates between services and components. Next, we identify the different types of services in Section \ref{sec:service-taxonomy}. The analysis of existing service identification approaches that consider service types into account is performed in Section \ref{sec:existing-identification-approaches}. In Section \ref{sec:propose-approach} and Section \ref{sec:conclusion}, we provide proposition to develop a service identification approach, and discuss the conclusion of this report, respectively.

\section{What is a Service?}
\label{sec:what-services}
In the literature, many definitions have been proposed for defining services \cite{barry2003web} \cite{nakamura2009extracting} \cite{erradi2006soaf} \cite{brown2002using} \cite{openGroup}. 
Each definition describes a service based on different details (e.g., granularity, communication mechanism, composition, etc.). In this section, we present some of these definitions.

\subsection*{Defining a simple service:}
Barry \cite{barry2003web} defined a service as \textquotedblleft \textit{a well-defined, self-contained function that does not depend on the context or state of other services}\textquotedblright  \cite{barry2003web}. However, this definition focuses on the functional attributes of the service, while it does not define service interfaces.

\subsection*{Considering service interfaces:}
Nakamura et al\footnote{A service is a set of processes that: (1) has an open interface. (2) self-contained (3) coarse-grained \cite{nakamura2009extracting}.} \cite{nakamura2009extracting} and Erradi et al\footnote{A service is a self-contained business functionality that has well-defined and discoverable interfaces \cite{erradi2006soaf}.} \cite{erradi2006soaf} added to these definitions the specification of service interfaces such that services should communicate through open and discoverable interfaces.

\subsection*{Additional issues:}
Some definitions, such as \cite{barry2003web} \cite{openGroup}, added more constraints to the service. For example, a service should be a black-box functionality based on the Open Group definition\footnote{A service is a black-box logical representation of a self-contained functionality that has a specified outcome and can be composed of other services \cite{openGroup}.} \cite{openGroup}. Some others allow a service to be composed of multiple other services to implement a larger functionality \cite{barry2003web} \cite{openGroup}. 

\subsection*{Definition of service interfaces:}
Brown et al\footnote{A service is: (1) composed of a loosely-coupled, coarse-grained, self-contained, discoverable, and composable functionality, (2) composed of multiple services that can be depends on each other, (3) communicates with other services based on clear interface via asynchronous messages \cite{barry2003web}.} \cite{brown2002using} clarified the communication model for service interfaces. This is based on a loosely coupled, (sometimes) asynchronous, message-based. 

\subsection*{Meaning of Service Characteristics}
\label{Terminologies}

\textbf{Open Interface:} "A service has an open interface, by which external entities can access to the service independently of the implementation of the service. For the access, a service cannot require platform specific operations, nor implementation-specific data that are only used within the system." \cite{nakamura2009extracting}

\textbf{Self-contained:} "A service can be executed by itself without any other services. Thus, a process cannot be a service if the process requires execution and/or data of any other processes. Such mutually-dependent processes should be aggregated within the same service." \cite{nakamura2009extracting}

\textbf{Coarse-Grained:} "A service is a coarse-grained process that can be a business construct by itself. Also, multiple services can be integrated to achieve a more sophisticated and coarser-grained service." \cite{nakamura2009extracting}

\textbf{Stateless} "means there is no record of previous interactions and each interaction request has to be handled based entirely on information that comes with it." [Internet dictionary]

\textbf{Discoverable} "means that services can be found at both design time and run time, not only by unique identity but also by interface identity and by service" \cite{adjoyanservice}.

\section{How are services different from components?}
\label{sec:service-vs-component}
Following the definitions of services \cite{barry2003web} \cite{nakamura2009extracting} \cite{erradi2006soaf} \cite{brown2002using} \cite{openGroup} and software components \cite{Baster2001BCC}\footnote{A component is \textquotedblleft abstract, self-contained packages of functionality performing a specific business function within a technology framework. These business components are reusable with well-defined interfaces\textquotedblright \cite{Baster2001BCC}.}, \cite{szyperski2002component}\footnote{A component is \textquotedblleft a unit of composition with contractually specified interfaces and explicit context dependencies only. A software component can be deployed independently and is subject to composition by third parties\textquotedblright \cite{szyperski2002component}.} \cite{luer2002composition}\footnote{A component is \textquotedblleft a software element that (a) encapsulates a reusable implementation of functionality, (b) can be composed without modification, and (c) adheres to a component model\textquotedblright \cite{luer2002composition}.}, we find that services are very similar to software components in terms of their characteristics (\textit{loose coupling}, \textit{reusability}, \textit{autonomy}, \textit{composability}, etc.). However, we can distinguish between services and components based on two aspects:

\subsection{The Granularity Range} 
\begin{figure}[h]
	\begin{center}
		\includegraphics[width=0.50\textwidth]{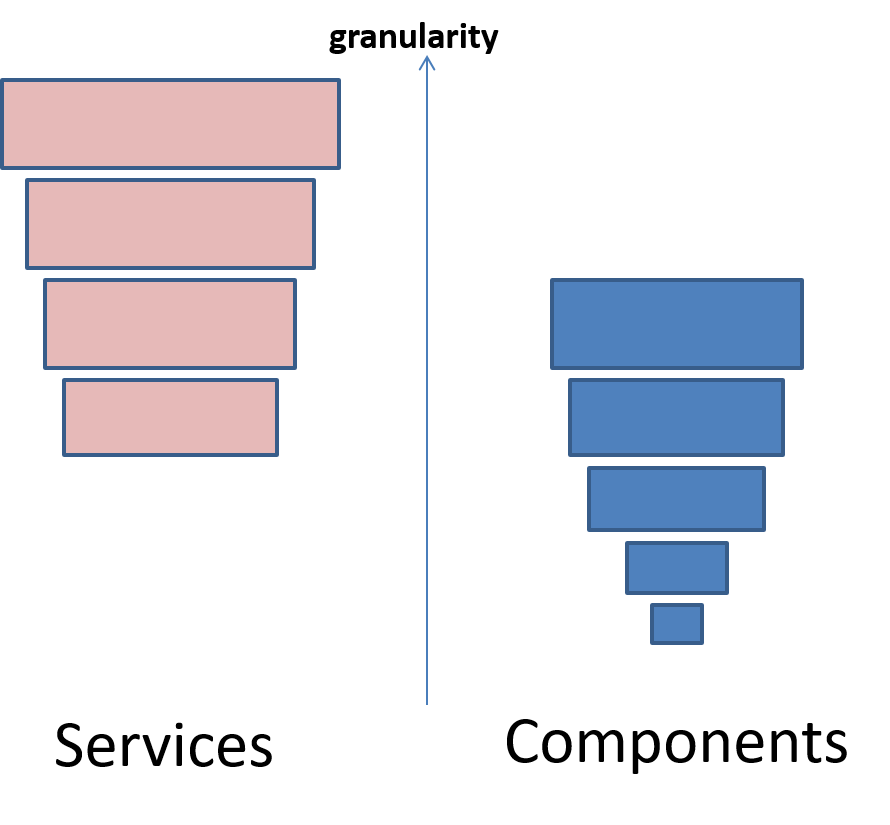}
		\caption{The granularity range of services and components}
		\label{fig:service-vs-component}
	\end{center}
\end{figure}

Services start at higher level of abstractions compared to components. A service can be a part of a business process (at the requirement level), an architectural element (at the design level) and a function (at the implementation level). However, components appear only at the design level and the implementation level in terms of architectural elements and functions (Figure \ref{fig:service-vs-component}.
).

\subsection{The Deployment Technologies and Models} 
Services and components are different in terms of deployment technologies and models that are used to technically implement them. For examples, services can be \textit{web-services} \cite{alonso2004web}, \textit{micro-services} \cite{namiot2014micro}, \textit{REST services} \cite{riva2009designing}, etc., while components can be \textit{OSGi} \cite{alliance2003osgi}, \textit{Fractal} \cite{bruneton2006fractal}, \textit{SOFA} \cite{plasil1998sofa}, etc. These have variations in their specification that make the implementation of services and components varied and respectively their provided and required interfaces.

\section{What are the Different Types of Services?}
\label{sec:service-taxonomy}
Several taxonomies \cite{alahmari2010service} \cite{fuhr2011using} \cite{gu2010service} \cite{grieger2014architectural} \cite{cohen2007ontology} \cite{dikmans2012soa} were presented to categorize different service types. In this section, we first present these six taxonomies. Then, we discuss our service taxonomy of service types as a result of the analysis of the existing ones.

\subsection{Existing  Taxonomies of Service Types}

\subsubsection{Alahmari's Service Taxonomy}
Alahmari et al. \cite{alahmari2010service} classified services into seven types based on their granularities (coarse-grained or fine-grained), purposes (CRUD) and the data they manipulate. These types are as follows:

\begin{itemize}
	\item \textbf{Process service:} is the largest coarse-grained service that performs a sequence of tasks corresponding to a business process. It can be composed of other types of services.
	\item \textbf{Business service:} implements a business logic/value.
	\item \textbf{Transactional-data service:} performs a CRUD functions (Create, Retrieve, Update, Delate) on transactional data.
	\item \textbf{Master-data service:} performs a CRUD functions on master data.
	\item \textbf{Utility service:} offers a domain functionality that is required by other services to perform their tasks.
	\item \textbf{Infrastructure service:} offers a technical functional service that is needed by other services.	
	\item \textbf{Composite service:} is the aggregation and orchestration of different atomic services.
\end{itemize}

\subsubsection{Fuhr's Service Taxonomy}
Fuhr et al. \cite{fuhr2011using} distinguished three types of services based on their relationships with the business process activities. These are:

\begin{itemize}
	\item \textbf{Business service:} provides the implementation of a specific business functionality corresponds only to one business activity.	
	\item \textbf{Utility service:} implements a functionality required by some other services. This type corresponds to services participating in different business activities.
	\item \textbf{Helper service:} implements a  general functionality that is required by most of the other services. Normally, it is used by all of the business activities.
\end{itemize}

\subsubsection{Qu's Service Taxonomy}
Qu and Lago \cite{gu2010service} presented six service types. This is based on studying 30 service identification approaches. These are:

\begin{itemize}
	\item \textbf{Business process service:} implements a business logic/value.
	
    \item \textbf{Data service:} concerns an entity object and/or a data one.
	
    \item \textbf{Composite service:} is the aggregation of other services.
	
    \item \textbf{IT service:} offers functionalities related to technology. This can be infrastructure or utility services.
	
    \item \textbf{Partner service:} is provided to an external partner.
    
	\item \textbf{Web service:} is implemented using Web-based technology.	
\end{itemize}

\subsubsection{Grieger's Service Taxonomy}
Grieger et al. \cite{grieger2014architectural} defined three types of services are identified, These are as follows:

\begin{itemize}
\item \textbf{Initial design service:} is a fine-grained functionality that is used to implement one single business logic/value.

\item \textbf{Composite service:} denotes to the building of a larger service based on aggregating other initial design service.

\item \textbf{Utility or technical service:} offers crosscutting domain and technical functionalities requires by other serviced.
\end{itemize}

\subsubsection{Cohen's Service Taxonomy}
Cohen \cite{cohen2007ontology} recognized six types of services that can be classified into two main categories based on whether a service is a part of the application implementation (i.e., application service) or a part of the platform (i.e., infrastructure service). Application services can be of four kinds: (1) entity, (2) activity, (3) capability, and (4) process services. Infrastructure services can be of two kinds: (5) communication and (6) utility services. These services are defined as follows:

\begin{itemize}
\item \textbf{Process service:} is the implementation of a business process by aggregating and orchestrating other types of services.	

\item \textbf{Capability service:} is the implementation of an action-centric functionality corresponds to a generic business logic/value. Its scope is organizational resource.

\item \textbf{Activity service:} is the implementation of an action-centric functionality corresponds to a generic business value. But, its scope is smaller than the capability service. It can be for a single application or a family composed of some applications.

\item \textbf{Entity service:} is a data-centric service corresponding to a business entity like employee or customer.

\item \textbf{Communication service:} is a server that has message transportation capabilities, regardless the message content.

\item \textbf{Utility service:} offers infrastructural functionality that is not tied to any specific business services.
\end{itemize}

\subsubsection{Dikmans's Service Taxonomy}
Dikmans \cite{dikmans2012soa} provided a comprehensive service categorization schema that is built based on six axes. These are related to the service granularity (elementary, composite and process), the actor who executes the service (human or IT-system), the channel used to offer the service (telephone, web, email, etc.), security level (public or confidential), organizational boundaries (department, enterprise or external), and the architecture layer (business, application or technical).

\begin{figure*}[!h]
	\begin{center}
		\includegraphics[width=\textwidth]{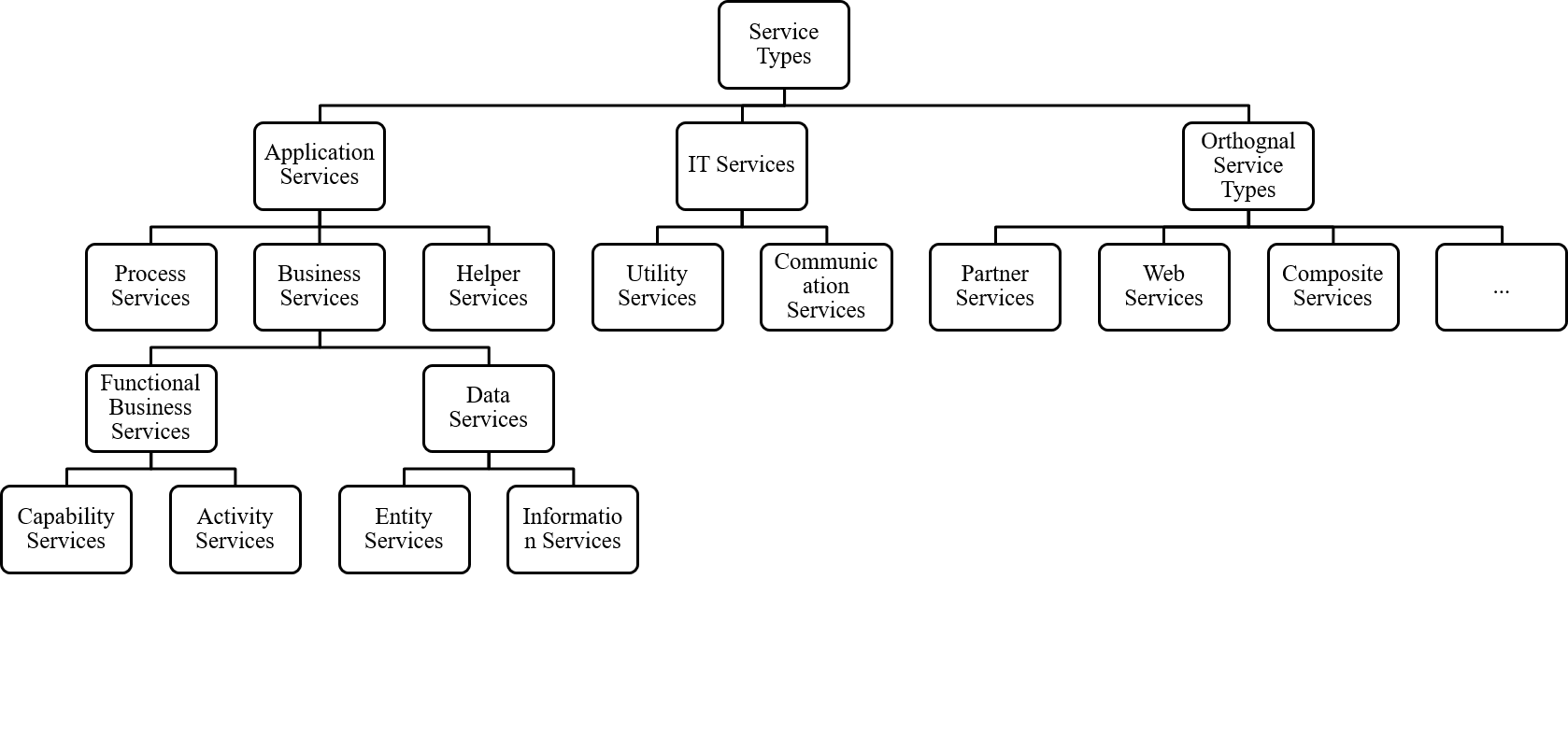}
		\caption{A synthesis of service type taxonomies}
		\label{fig:servicetypes}
	\end{center}
\end{figure*}

\subsection{Synthesis of Our Service Taxonomy}

In the previous subsection, we presented six service taxonomies that distinguish different service types. These taxonomies are similar in terms of some service's types and differ in terms of some others. In this section, we want to provide a comprehensive view of these taxonomies and build our taxonomy that covers all of them. 
Figure \ref{fig:servicetypes} shows our taxonomy. We classify all service mainly into two types of services based on application and IT perspectives. In the following subsections, we will discuss them in details.

\subsubsection{Applications services}
Application services implement functionalities related to business values in the application implementation space. We classify them based on their granularities into three categories:
\begin{enumerate}
	\item \textbf{Process services} implement largest coarse-grained functionalities correspond to business processes. This type of services is only considered in the taxonomies of Alahmari et al. \cite{alahmari2010service} Cohen \cite{cohen2007ontology} and Dikmans \cite{dikmans2012soa}. 
	
	\item \textbf{Business services} implement specific coarse-grained business functionalities correspond to business activities or tasks. This type is defined by all of the mentioned taxonomies. Some researchers distinguished different business services. Cohen \cite{cohen2007ontology} and Qu \cite{gu2010service} classified business services into two kinds. The first one refers to data services that implement entities and data related services. The second kind denotes to business services that implement functionalities related to business logics/values.  The latter is also classified by Cohen \cite{cohen2007ontology} into two types based on the service scope. A service is called capability service when it has a large scope such as organization scope, while it is called activity service when it has a small scope such as a single application or a single family composed of few applications (i.e., a service in a software product line). 
	
	\item \textbf{Helper services} that implements fine-grained general functionalities related to small business values shared by different business logics. This means that these services are used by other application services to perform their activities. Helper services are referred by Alahmari et al. \cite{alahmari2010service}, Grieger et al. \cite{grieger2014architectural} and Fuhr et al. \cite{fuhr2011using}. For example, a helper service provides CRUD functions for another business service \cite{alahmari2010service}. Cohen \cite{cohen2007ontology} and Qu et al. \cite{gu2010service} do not consider this type of services.

\end{enumerate}

\subsubsection{IT services}

IT services provide technical functionalities offered by the infrastructure platforms and operating systems. These services are provided through programming languages API (e.g., \textit{Java SDK}, \textit{Android SDK}, \textit{.Net Framework}, etc.). Application services require them to perform their tasks. This type is used by Alahmari et al. \cite{alahmari2010service}, Qu et al. \cite{gu2010service}, and Grieger et al. \cite{grieger2014architectural} and Cohen \cite{cohen2007ontology}. Cohen \cite{cohen2007ontology} considered additional sub-type of IT services. This refers to communication services that only provide message transportation capabilities, regardless the message content. IT services are not considered by Fuhr et al. \cite{fuhr2011using} taxonomy.
	
\subsubsection{Orthogonal Service Types}
Other orthogonal service types are distinguished in the service taxonomies. We consider these types of services as attributes that can be applied to any of the previously mentioned services regardless their types. 

Qu et al. \cite{gu2010service} considered a service as a partner service if it offers interfaces to external software, and a service is considered as a web service if it is implemented using a Web technology. 

We consider composite service type as a methodology to build a service/application by orchestrating different services, in order to produce a larger service that implements a larger business value.
For example, we orchestrate several business services and infrastructure services to form a single process service related to a business process. Composite services are presented in Alahmari et al. \cite{alahmari2010service}, Qu et al \cite{gu2010service}, Grieger et al. \cite{grieger2014architectural}, and Cohen \cite{cohen2007ontology} taxonomies. 

Many other orthogonal service types are presented by Dikmans \cite{dikmans2012soa}. These types are based on the actor (human or IT-system), the channel used (telephone, web, email, etc.), security level (public or confidential), organizational boundaries (department, enterprise or external), and the architecture layer (business, application or technical).

\section{What are the Existing Service Identification Approaches that Consider Service Types in to Account?}
\label{sec:existing-identification-approaches}
In the literature, several approaches were presented to identify services from legacy software like \cite{nakamura2009extracting} \cite{alahmari2010service} \cite{fuhr2011using} \cite{grieger2014architectural} \cite{adjoyanservice} \cite{zhang2005service}.
However, we identify only three approaches \cite{alahmari2010service} \cite{fuhr2011using} \cite{grieger2014architectural} that consider different service types into account during the service identification process. In the following subsections, we summarize these approaches.

\subsection{Alahmari's Service Identification Approach}
Alahmari et al. \cite{alahmari2010service} identified services based on analyzing business process models. These business process models are derived from questionnaires, interviews and available documentations that provide atomic business processes and entities on the one hand, and activity diagrams that provide primitive functionalities  on the other hand. The  activity diagrams are manually identified from UML class diagrams extracted from the legacy code using IBM Rational Rose.
Different service granularity are distinguished in relation to atomic business processes and entities. Dependent atomic processes as well as the related entities are grouped together at the same service to maximize the cohesion and minimize the coupling. There is no details about how to identify the different service types.

\subsection{Fuhr's Service Identification Approach}
In \cite{fuhr2011using}, Fuhr et al. identified three types of services. These are business, utility and helper services. The services are identified from legacy codes based on a dynamic analysis technique. The authors relied on a business process model to identify correlation among classes. Each activity in the business process model is executed. Classes that have got called during the execution are considered as related.
The identification of services is based on a clustering technique where the similarity measurement is based on how many classes are used together in the activity executions. The identified clusters are manually interpreted and mapped into the different service types. Classes used only for the implementation of one activity are grouped into a business service corresponding to this activity. Utility services are composed of clusters of classes that contribute to implement multiple activities but not all of them. A Cluster of classes that are used by all of the activities represent the implementation of helper services. A strong assumption regarding this approach is that business process model should be available to identify execution scenarios.

\subsection{Grieger's Service Identification Approach}
Grieger et al. \cite{grieger2014architectural} presented an approach that identified three service types based on analyzing existing legacy modules. 
The first one refers to initial design services that implement business values. These are identified based on refining the existing legacy modules related to business values.
The second one denotes to coarse-grained services, e.g., related to business processes. These are identified based on orchestrating other services related to the same underlying business process (i.e., structural dependent services). To this end, a hierarchical clustering algorithm is utilized to identify a dendrogram tree that represents the aggregation of initial design services based on their dependencies.
The last service type is related to services that implement crosscutting concerns and technical functionalities used across different services (i.e., utility or technical services). The identification of these services is based on partitioning the functionalities of multiple services to recover individual and common parts. To this end, the authors relied on a clone detection algorithm to extract cloned functionalities shared among different services. Then, the identified cloned functionalities are given to software architects to decide if these cloned functionalities need to be moved into an existing service or to create a new one.

\section{How to Identify Services Based on the Object-Oriented Source Code with Respect to their Types?}
\label{sec:propose-approach}
In this section, we provide an overview analysis of how to propose a service identification approach that identify different service types from legacy object-oriented software based on the analysis source code.
We identify three main elements that have to be defined for any service identification approach. These are:

\begin{enumerate}
	\item \textbf{Service Compared to Object:} we want to to explain what do services structurally mean compared to objects. Thus, we map service-oriented concepts to object-oriented concepts.
	
	\item \textbf{Target Service Types:} service types mentioned previously by service taxonomies can not be always identified only based on the source code  analysis. Some types require additional software artifacts to be analyzed. Thus, we discuss what types of services that we are interested (based on the possibilities to identify them) to identify based on the source code analysis.
	
	\item \textbf{Identification Algorithms:} the identification approach can be applied based on different algorithms to identify different service types. This is based on where service types are differentiated compared to the service identification step. Thus, we provide our proposition about how these algorithms work.

\end{enumerate}

\subsection{Service Compared to Object}
In object-oriented software, functionalities are implemented in source codes in terms of a set of object-oriented classes organized in a set of object-oriented packages. Each class encapsulates a set of methods (operations) and attributes (data) related to one object.

We structurally define a service as a collection of object-oriented classes that participate to implement a specific set of related functionalities. The service's functionalities are accessible by other services through the public methods that are encapsulated in the object-oriented implementation of the service. Thus, the service interfaces are structured based on the set of all public methods existed in the object-oriented classes composing the implementation of a service.

\subsection{Target Service Types}
In the context of service reverse engineering, distinguishing between different service types cannot be always identified based on the analysis of source codes. Some service types require further investigation based on additional software artifacts. 

The identification step of some service types applied the same process, while the difference between their types is based on the software architect knowledge. For example, the difference between capability services and activity services requires to identify the service scope, which is not available in the source codes. This requires to recover where the service can be deployed, in a single system or in the organization\footnote{Can be related to software product lines!}. These are identified as business services with different attributes.

We focus on the following three service types during the design of service identification approaches:

\textbf{Process service type} is used to implement a large-grained service related to business process.  This business process is composed of a set of business activities that are implemented based on a set of business services. These business services depend on other helper services as well as IT ones. Thus, we invest the service composition type to define process service type. Therefore, a process service can be identified as a composition of a collection of dependent business, helper and IT services that are participate to implement the corresponding business process.

\textbf{Business service type} is used to realize the implementation of coarse-grained services corresponding to business activities. Thus, we define a business service as a specific coarse-grained functionality that provides the implementation of one business activity.

\textbf{Helper service type} refers to services that implement functionalities that do not have business values. However, they are used by business services to perform their tasks. For example, non-functional functionalities.

\textbf{IT service type} is used to implement services related to technical infrastructural services that are related to the platforms. These are used by other service types to access the platform capabilities through programming languages.

\subsection{Identification Algorithms}
\label{section:IdentificationAlgorithm}

\begin{figure*}[]
	\begin{center}
		\includegraphics[width=\textwidth]{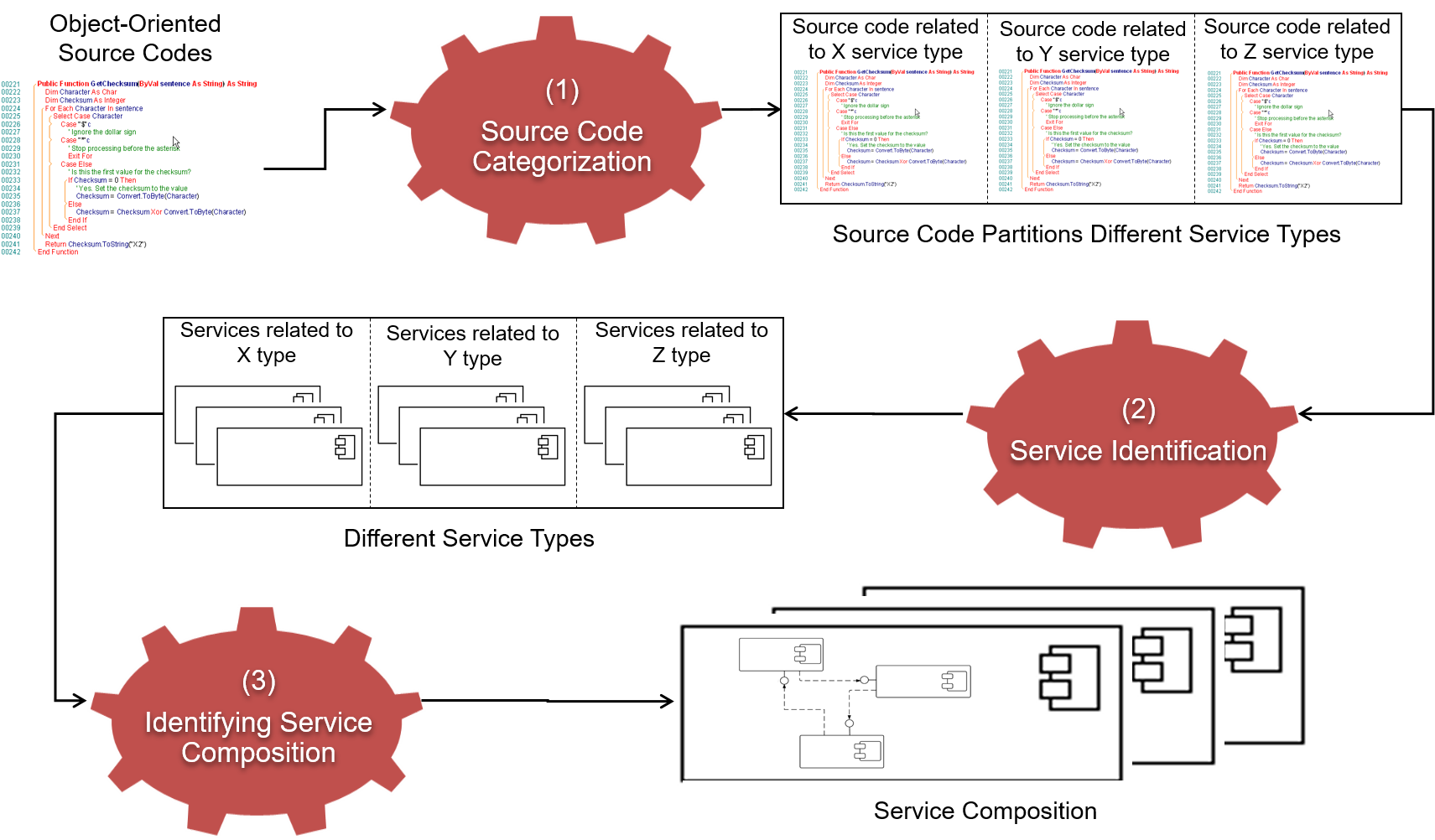}
		\caption{Process of pre-identification algorithm}
		\label{fig:processPreIdenti}
	\end{center}
\end{figure*}

\begin{figure*}[]
	\begin{center}
		\includegraphics[width=\textwidth]{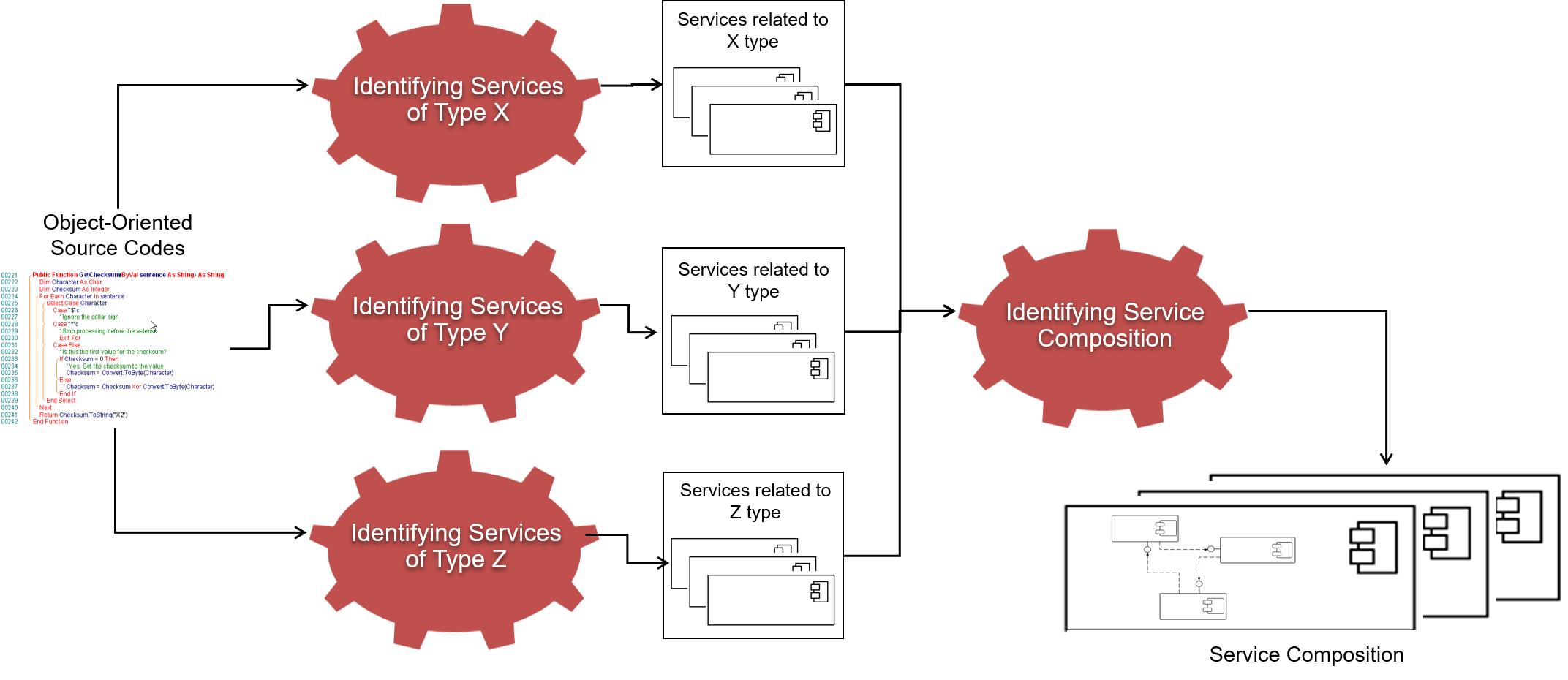}
		\caption{Process of in-identification algorithm}
		\label{fig:processInIdenti}
	\end{center}
\end{figure*}

\begin{figure*}[]
	\begin{center}
		\includegraphics[width=\textwidth]{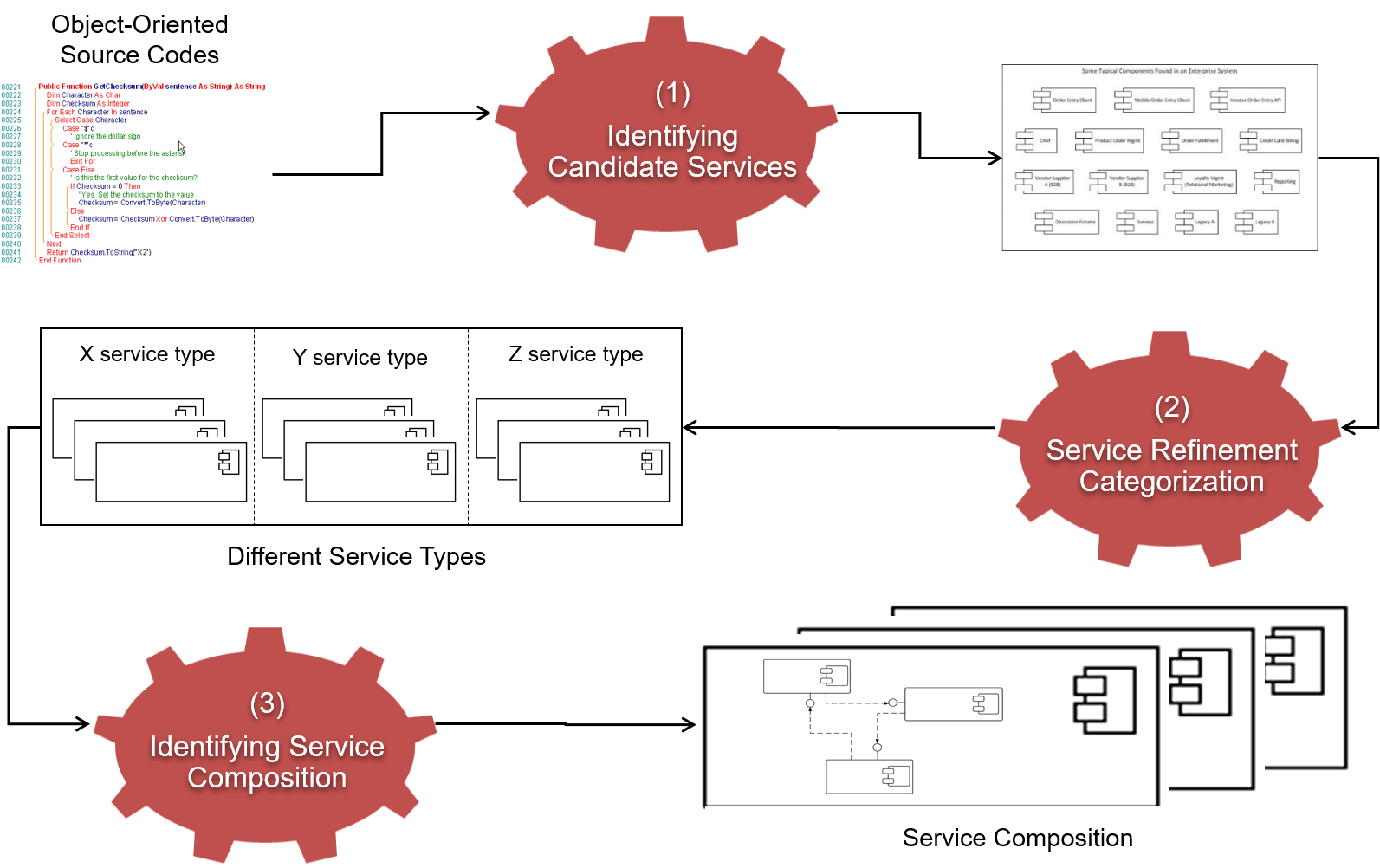}
		\caption{Process of post-identification algorithm}
		\label{fig:processPostIdenti}
	\end{center}
\end{figure*}

We distinguish three service identification algorithms based on where service types are taken into account in relation to the service identification method. These are pre-identification, in-identification and post-identification algorithms.
The pre-identification algorithm concerns the idea of partitioning object-oriented source code into different parts based on the target service types. Then, a service identification method is applied on each part to recover the corresponding services (see Fig. \ref{fig:processPreIdenti}).
The in-identification one is related to design several identification algorithms where each algorithm is interested at identifying a specific service type based on the analysis of source codes (see Fig. \ref{fig:processInIdenti}).
The post-identification algorithm is related to one service identification process followed by refinement and classification processes to differentiate service types (see Fig. \ref{fig:processPostIdenti}). 

\section{Conclusion}
\label{sec:conclusion}
In this report, we discussed a set of issues that help developing a service identification approach based on the analysis of object-oriented source of legacy software. Such issues answer a set of research questions that are: 

\begin{enumerate}
\item \textbf{What is a service?} We analyzed existing service definitions. We organized these definitions based on the details that they provide about services.

\item \textbf{How are services different from components?} We clarified how services are different from software components based on the granularity range and the deployment technologies and models.

\item \textbf{What are the different types of services?} We analyze six service taxonomies that were presented to classify different service types. Also, we discussed a new service taxonomy of service types based on the analysis of the existing ones.

\item \textbf{What are the existing service identification approaches that consider service types into account?} 
We presented three service identification approaches that considered different service types into account during the service identification process.

\item \textbf{How to identify services based on the object-oriented source code with respect to their types?}
We showed the mapping of service-oriented concepts to object-oriented concepts. Then, we identify the set of service types that can be identified based on the source code analysis. Next, we discussed the algorithm that can be used to identify services and distinguishing their types.

\end{enumerate}

\bibliographystyle{unsrt}
\bibliography{main}  % sigproc.bib is the name of the Bibliography in this case

\noindent
\end{document}